\documentclass[aps,prl,preprint,groupedaddress,floatfix]{revtex4}
\usepackage{graphicx}
\begin{document}

\title{
$\beta$ decay of neutron-rich $^{53-56}$Ca
}

\author{P.F.~Mantica$^{(1,2)}$} 
\author{R.~Broda$^{(3)}$}
\author{H.L.~Crawford$^{(1,2)}$}
\author{A.~Damaske$^{(1,2)}$}
\author{B.~Fornal$^{(3)}$}
\author{A.A.~Hecht$^{(4,5)}$} 
\author{C.~Hoffman$^{(6)}$}
\author{M.~Horoi$^{(7)}$}
\author{N.~Hoteling$^{(4,5)}$}
\author{R.V.F.~Janssens$^{(5)}$}
\author{J.~Pereira$^{(1,8)}$ } 
\author{J.S.~Pinter$^{(1,2)}$}
\author{J.B.~Stoker$^{(1,2)}$}
\author{S.L.~Tabor$^{(6)}$}
\author{T.~Sumikama$^{(9)}$}
\thanks{Present address: Department of Physics,
Tokyo University of Science, 2641 Yamazaki, Noda, Chiba
278-8510, Japan.}
\author{W.B.~Walters$^{(6)}$} 
\author{X.~Wang$^{(5,10)}$}
\author{S.~Zhu$^{(5)}$}

\affiliation{$^{(1)}$ 
National Superconducting Cyclotron
Laboratory, Michigan State University,
East Lansing, Michigan 48824, USA}
\affiliation{$^{(2)}$
Department of Chemistry, Michigan State University,
East Lansing, Michigan 48824, USA}
\affiliation{$^{(3)}$
Institute of Nuclear Physics, Polish Academy of Sciences,
PL-31342, Cracow, Poland}
\affiliation{$^{(4)}$
Department of Chemistry and Biochemistry, University 
of Maryland, College Park, Maryland 20742, USA} 
\affiliation{$^{(5)}$ 
Physics Division,
Argonne National Laboratory, Argonne, Illinois 60439, USA}
\affiliation{$^{(6)}$
Department of Physics and Astronomy, Florida State University,
Tallahassee, Florida 32306, USA}
\affiliation{$^{(7)}$
Department of Physics, Central Michigan University,
Mount Pleasant, Michigan 48859, USA}
\affiliation{$^{(8)}$
Joint Institute for Nuclear Astrophysics, Michigan State University,
East Lansing, Michigan 48824, USA}
\affiliation{$^{(9)}$
RIKEN, 2-1 Hirosawa, 
Wako-shi, Saitama 351-0198, Japan}
\affiliation{$^{(10)}$
Department of Physics, University of Notre Dame,
South Bend, Indiana 46556, USA}

\date{\today}

\begin{abstract}

$\beta$-decay properties of neutron-rich Ca isotopes have been 
obtained.  Half-life values were determined for the 
first time for $^{54}$Ca ($86\pm 7$~ms), 
$^{55}$Ca ($22\pm 2$~ms), and $^{56}$Ca ($11\pm 2$~ms).
The half-life of $230\pm 60$~ms deduced for $^{53}$Ca
is significantly longer than reported previously,
where the decay chain $^{53}$K $\rightarrow$ $^{53}$Ca $\rightarrow$
$^{53}$Sc was considered. A delayed $\gamma$ ray with
energy 247~keV was identified following $\beta$ decay of 
$^{54}$Ca, and is proposed to depopulate the $1^{+}_{1}$ level
in $^{54}$Sc.  The $\beta$-decay properties compare
favorably with the results of shell model calculations completed in the 
full $pf$-space with the GXPF1 interaction.  The half-lives of 
the neutron-rich Ca isotopes are also compared with 
gross $\beta$-decay theory.  The systematic trend of the 
neutron-rich Ca half-lives is consistent with the presence of a 
subshell gap at $N=32$.
\end{abstract}
\maketitle

\section{Introduction}

The appearance of added stability in atomic nuclei 
at certain nucleon numbers prompted the development 
of the nuclear shell model.  The robustness of the magic numbers
at 2, 8, 20, 28, 50, 82, and 126 has been well 
demonstrated for nuclei near the valley of $\beta$
stability.  However, only limited data are available
for $\beta$-unstable nuclei with magic numbers 
of protons and/or neutrons.  In several instances, 
the observed properties of such radioactive nuclei
are not consistent with predictions assuming 
stabilized proton and/or neutron cores.  
As examples, the non-existence of 
$^{28}_{8}$O$_{20}$ \cite{tar1997},
the appearance of an island of inversion around
$^{32}$Mg$_{20}$ \cite{war1975,mot1995} and
suspected collectivity at $^{56}_{28}$Ni$_{28}$ \cite{yur2004,min2006} 
all call to question a simple extrapolation
of the known magic numbers to the drip-lines of the nuclear 
chart.

The Ca isotopes have a closed-shell number of 
protons ($Z=20$) and have been well studied in the range
of the doubly-magic nuclei $^{40}$Ca$_{20}$
and $^{48}$Ca$_{28}$, where neutrons fill the $1f_{7/2}$ 
orbital.  
Beyond doubly-magic $^{48}$Ca, neutrons begin
filling the upper $pf$ shell.  
Neutron transfer data \cite{can1971,tac1978,uoz1994} are available 
for levels in $^{49}$Ca$_{29}$, and the deduced spectroscopic factors 
signify energy gaps between the adjacent
$2p_{3/2}$, $2p_{1/2}$, and $1f_{5/2}$ single-particle orbitals.
Such gaps suggest added stability for the 
Ca isotopes with $N = 32$ and $N = 34$.
Indeed, a subshell closure at $N = 32$
for $^{52}$Ca \cite{huc1985,gad2006}, $^{54}$Ti \cite{jan2002},
and $^{56}$Cr \cite{pri2001} has been inferred from
the systematic variation of first-excited $2^+$
states in the even-even isotopes of these 
elements as a function of 
neutron number.  
The yrast structures of the even-even Cr \cite{app2003,zhu2006} 
and Ti \cite{jan2002,for2004} isotopes at higher spins 
also show irregular
energy spacings at $N=32$ that can be understood if a 
significant gap in neutron single-particle states between
$\nu p_{3/2}$ and $\nu p_{1/2}$, $\nu f_{5/2}$ exists.  
Additional support for the apparent shell gap at $N=32$ is found
in the small $B(E2:0^+ \rightarrow 2^+)$ values measured 
for $^{54}$Ti$_{32}$ \cite{din2005} and $^{56}$Cr$_{32}$ \cite{bur2005}.
No substantial experimental evidence for an 
expected \cite{hon2002} subshell closure at 
$N=34$ has been found to date, as the 
even-even Cr \cite{pri2001,sor2003} and Ti \cite{lid2004-prl}
isotopes show a decrease in energy of the $2^+_1$ level
when moving from $N=32$ to $N=34$.  

Little data are available on the structures of the neutron-rich
Ca isotopes beyond $^{52}$Ca$_{32}$.  Knowledge of existence has
only been extended to $^{56}$Ca$_{36}$ \cite{ber1997}.
The heaviest Ca isotope where structure and half-life information
is available is $^{53}$Ca.  The $\beta$-decay 
half-life of $^{53}$Ca was deduced to be $90 \pm 15$~ms \cite{lan1983}.
In that study, $^{53}$Ca was produced from the
decay of $^{53}$K,  which was made directly by fragmentation of 
an Ir target by light ions, and extracted from an ion source.  
The mass-separated sample was 
studied via $\beta$-delayed neutron spectroscopy.  A two-component
decay curve was observed, and the component with the longer
decay constant was attributed to the daughter $^{53}$Ca decay.
A delayed neutron branch of ($40 \pm 10$)\% was obtained 
for $^{53}$Ca.  A subsequent $\beta$-decay study of $^{53}$K  
\cite{per2006} revealed a single bound state at 2.2~MeV above
the $^{53}$Ca ground state.  Significant neutron branching was 
also observed for the decay of $^{53}$K, but no additional information
on the $\beta$-decay properties of the $^{53}$Ca daughter was
reported in Ref. \cite{per2006}.
 
We report results of $\beta$-delayed $\gamma$-ray 
spectroscopic measurements of $^{53-56}$Ca to examine the robustness
of the $Z = 20$ proton shell closure for very
neutron-rich Ca isotopes.  The half-lives of 
$^{54-56}$Ca were determined for the first time,
and a longer half-life value was deduced for $^{53}$Ca.  
The systematic trend in half-life values for the 
neutron-rich Ca isotopes is compared to predictions 
from gross $\beta$ decay and shell model calculations completed
in the full $pf$-shell model basis.  

\section{Experimental Methods}

\begin{figure}[h]
\includegraphics{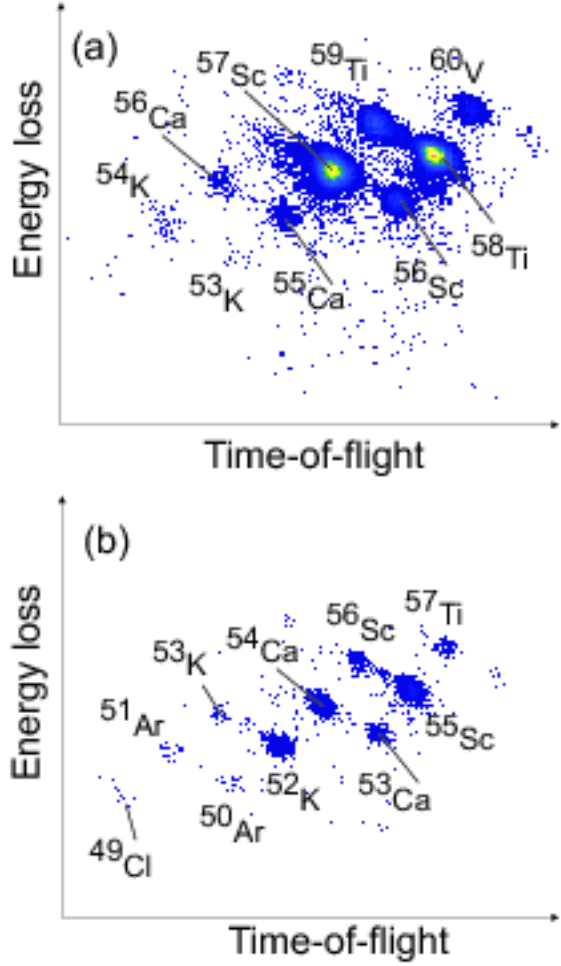}
\caption{(Color online) Particle identification spectra for 
A1900 spectrometer settings where (a) $^{54}$K and (b) $^{54}$Ca
were centered in the dispersive plane at the final focus of the device.
The energy loss measurement is from the most upstream Si PIN detector
of the Beta Counting System with thickness 991 $\mu$m.}
\label{fig1}
\end{figure}

Neutron-rich Ca nuclei were produced by 
fragmentation of a 140 MeV/nucleon $^{76}$Ge 
beam at National Superconducting Cyclotron 
Laboratory at Michigan State University.
The primary beam was incident on a 352 mg/cm$^{2}$ 
Be target at the object position of the A1900 
fragment separator \cite{mor2003}.  
Two independent settings of the A1900 separator 
were used to separate the desired isotopes from
unwanted reaction products.  The first 
setting was optimized for production of 
$^{54}$K with magnetic rigidities
B$\rho _1 = 4.4030$~Tm and B$\rho _2 = 4.1339$~Tm, and
included $^{55,56}$Ca.  
The second A1900 setting with B$\rho _1 = 4.3867$~Tm 
and B$\rho _2 = 4.1339$~Tm was chosen to center the production 
of $^{54}$Ca at the A1900 focal plane without changing the 
tune of the second half of the spectrometer (and the downstream
beam transport magnets).  The particle identification 
spectra collected at the experimental end station for 
both A1900 settings are presented in Fig.\ \ref{fig1}.
The full momentum acceptance of the A1900 ($\sim 5$\%) was used 
for fragment collection.  An Al wedge of thickness 300 mg/cm$^{2}$ 
was placed at the intermediate 
momentum-dispersive image to provide differential energy loss 
to fragments of differing $Z$ traversing the spectrometer.  
A thin plastic scintillator detector was also placed near this position,
and position information deduced from this detector was used
to correct the fragment time of flight.      

The experimental end station consisted of detectors 
from the Beta Counting System \cite{pri2003} and 
the Segmented Germanium Array \cite{mue2001}.  
The fragments transported from the A1900 to the 
experimental end station first encountered a stack
of three Si PIN detectors with thicknesses 991, 997, and
309 $\mu$m.  These detectors provided energy loss
information on the incoming beam, critical for 
particle identification.  The timing signal from 
the most upstream Si PIN was used in conjunction with
the cyclotron radiofrequency to deduce fragment 
time of flight information.  The Si PIN stack was 
followed by a Si double-sided multistrip
detector (DSSD) and six Si single-sided multistrip detectors
(SSSDs). The DSSD had thickness 979 $\mu$m and was segmented
into 40 strips on both front and back, for a total
of 1600 pixels.  The SSSDs had
nominal thicknesses of 1~mm and were segmented into 16 strips.
The SSSDs strip orientation alternated 
between horizontal and vertical (relative to ground) 
for each successive detector to provide two-dimensional 
position information for particles moving downstream
of the DSSD.  The majority of fragments were stopped 
in either the DSSD or the most upstream SSSD; a consequence
of the large momentum acceptance of the A1900 
separator.  

Sixteen detectors from the Segmented 
Germanium Array were 
arranged in two concentric rings around the 
vacuum beam line surrounding the Si detectors
of the Beta Counting System.  
The $\gamma$-ray peak detection efficiency was 
20\% at 100~keV and 7\% at 1~MeV.  
The energy resolution of each Ge detector
in the array was $\sim 3.5$~keV for 
the 1.3~MeV $\gamma$-ray transition in $^{60}$Co.

The master trigger for 
event readout was determined by a logical OR 
of any of the following three conditions:
1) signal above threshold in any of the 
16 strips of the most upstream 
SSSD; 2) signal above threshold 
in any of the 40 front strips AND
any of the 40 back strips of the DSSD;
3) signal above threshold in the most
upstream Si PIN detector (used for 
prompt $\gamma$-ray detection).
$\gamma$ rays detected within 15~$\mu$s of 
a master trigger were recorded into the event
stream.  The long gating time for $\gamma$ rays 
enabled identification of potential microsecond isomers 
populated either in the fragmentation process 
or following $\beta$ decay.  
Each event was tagged with an 
absolute time stamp based on a 50~MHz oscillator 
that was input to a 32-bit free-running scaler. 
Details of an implantation event, including 
energy loss and time-of-flight information 
required for particle identification, 
were stored in a two-dimensional array with indices
based on pixel location.  
Correlations with the energy-loss signal
from $\beta$-decay electrons detected in the DSSD were
then processed, based on 
temporal and position information.
Position correlations were restricted to 
$\beta$-decay electrons that emanated from the 
same pixel or one of the surrounding eight
pixels of an implantation. Time correlations
had two defined limits: a maximum  
time for a single $\beta$-decay event for a 
given implantation event and a minimum time 
for back-to-back implantation events into the 
same pixel.  The details of the implantation
event were cleared from the two-dimensional 
array when these defined limits were not satisfied. 
The results presented in the 
next section were obtained with a 
maximum correlation time of $1$~s and a 
minimum time for back-to-back implantations 
of $1$~s.

\section{Results}

\begin{figure}[h]
\includegraphics{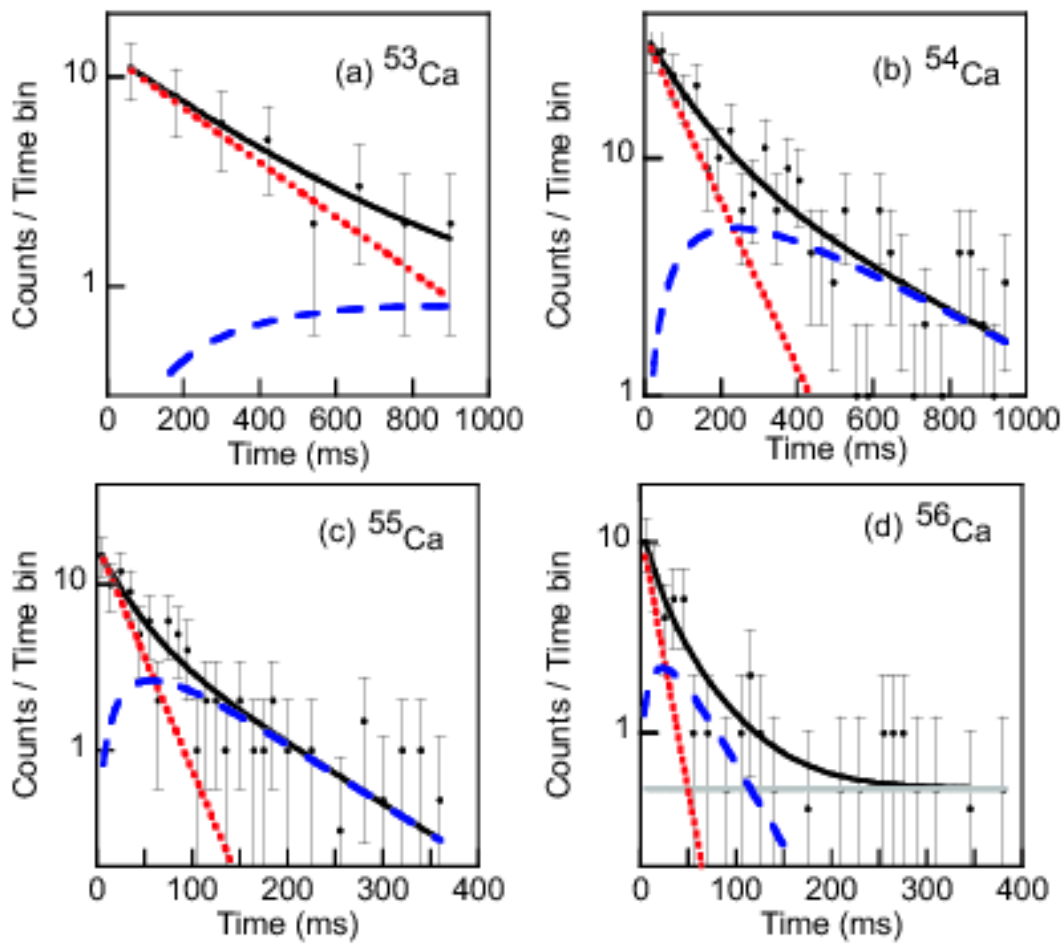}
\caption{(Color online) Decay curves for: (a) $^{53}$Ca; (b) $^{54}$Ca;
(c) $^{55}$Ca; and (d) $^{56}$Ca.  The fitting function (solid black line)
considered contributions from parent exponential decay (dotted line), daughter
exponential growth and decay (dashed line), and a flat background (solid gray line).  
The only fitted curve that resulted in a non-negligible value for the 
background term was that for $^{56}$Ca.  Additional details regarding the 
fitting procedure used to extract half-life values are 
provided in the text.}
\label{fig2}
\end{figure}

The decay curves for $^{53-56}$Ca are given 
in Fig.\ \ref{fig2}.  The half-life of 
each nuclide was deduced by fitting the 
decay curves to a function that considered 
exponential decay of the parent, exponential 
growth and decay of the daughter, 
and constant background.  
Decay information on the daughter 
nuclide was taken from the literature,
with additional details provided in the following
subsections. Table \ref{tab1} summarizes the 
half-life results for $^{53-56}$Ca.

\begin{table}[h]
\caption{Correlation statistics and half-lives deduced for the 
neutron-rich $^{53-56}$Ca isotopes.} 
\label{tab1}
\begin{ruledtabular}
\begin{tabular}{ccccc}
&\multicolumn{2}{c}{No.\ Parent Nuclei}&\multicolumn{2}{c}{$T_{1/2}$ (ms)}\\
\cline{2-3} \cline{4-5} \\
Nuclide&Implanted&Correlated&This Work&Previous \\
$^{53}$Ca  & 209 &  32 & $230 \pm 60$ &  $90 \pm 15$ \footnotemark[1]\\
$^{54}$Ca  & 654 & 136 & $ 86 \pm  7$ &                              \\
$^{55}$Ca  & 246 &  52 & $ 22 \pm  2$ &                              \\
$^{56}$Ca  &  99 &  18 & $ 11 \pm  2$ &                              \\
\end{tabular}
\end{ruledtabular}
\footnotetext[1]{Ref.\ \protect\cite{lan1983}}.
\end{table}

\begin{figure}[h]
\includegraphics{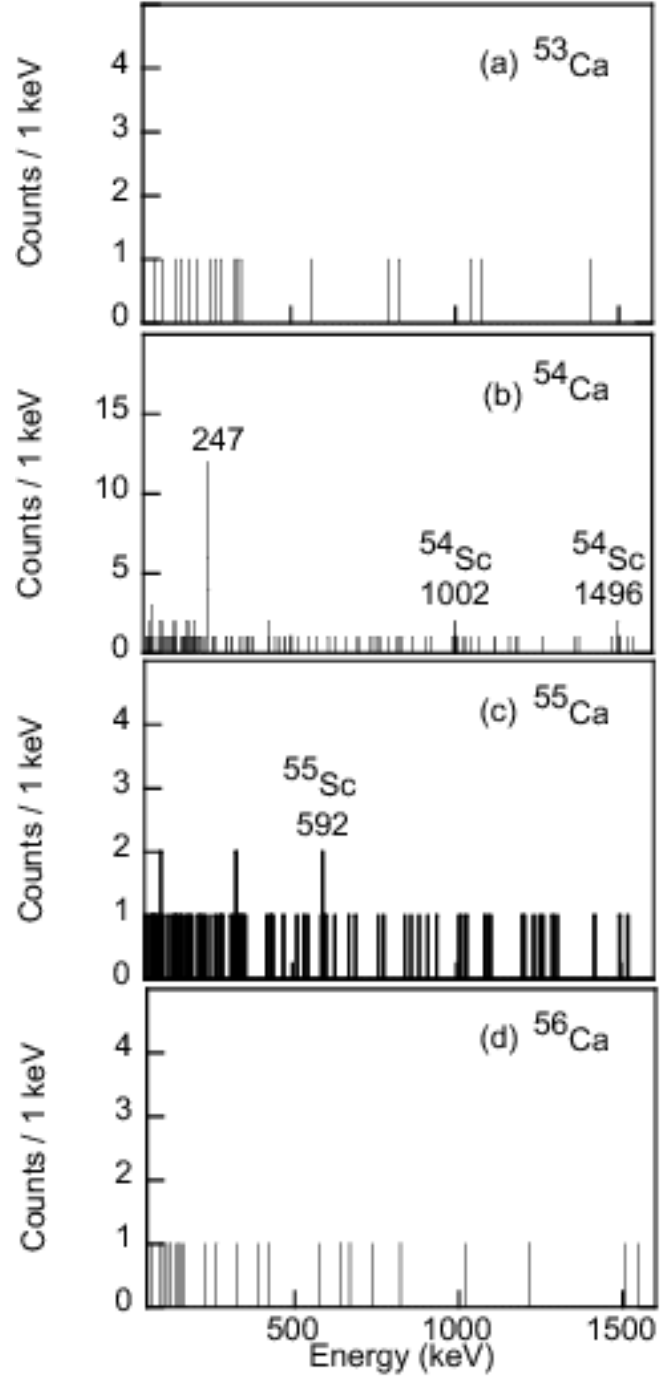}
\caption{$\gamma$-ray spectrum in the 
range 50 to 1600~keV correlated 
with $\beta$-decay events for: (a) $^{53}$Ca; (b) $^{54}$Ca;
(c) $^{55}$Ca; and (d) $^{56}$Ca.  Observed transitions 
are marked by their energy in keV.}
\label{fig3}
\end{figure}

The spectra of $\gamma$ rays correlated 
with the $\beta$ decays of $^{53-56}$Ca 
are presented in Fig.\ \ref{fig3}.  $^{54}$Ca 
is the only nuclide where a delayed $\gamma$ ray is 
assigned to the parent $\beta$ decay. 
No evidence for isomeric $\gamma$-ray emission within
15~$\mu$s of an implantation was found for the 
neutron-rich $^{53-56}$Ca isotopes.  

More detailed discussions are included below 
on the results for each isotope.

\subsection{$^{53}$Ca}

The half-life of $^{53}$Ca reported here, $T_{1/2} = 230 \pm 60$~ms, is significantly 
longer than the value $90 \pm 15$~ms reported in Ref.\ \cite{lan1983}.
The previous half-life value was deduced following the decay chain
$^{53}$K $\rightarrow$ $^{53}$Ca $\rightarrow$ $^{53}$Sc, where
$^{53}$K was directly produced by proton spallation of an Ir target
and only $\beta$ particles in coincidence with delayed neutrons were 
recorded.  In that work, data collected in the range 300 to 700~ms were fitted with a 
single exponential decay, assuming contributions from neither the 
parent nor the grand-daughter over this time range.  These assumptions
are reasonable considering the short half-life of $^{53}$K, 
$30 \pm 5$~ms \cite{lan1983}, and the unknown, but expected long 
($> 3$~s) half-life of $^{53}$Sc \cite{sor1998}.  
The decay curve presented in Fig.\ \ref{fig2} 
considers only direct production of $^{53}$Ca, with no ambiguity regarding 
the fit to a convoluted spectrum involving both parent and daughter 
decays.  

\begin{figure}[h]
\includegraphics{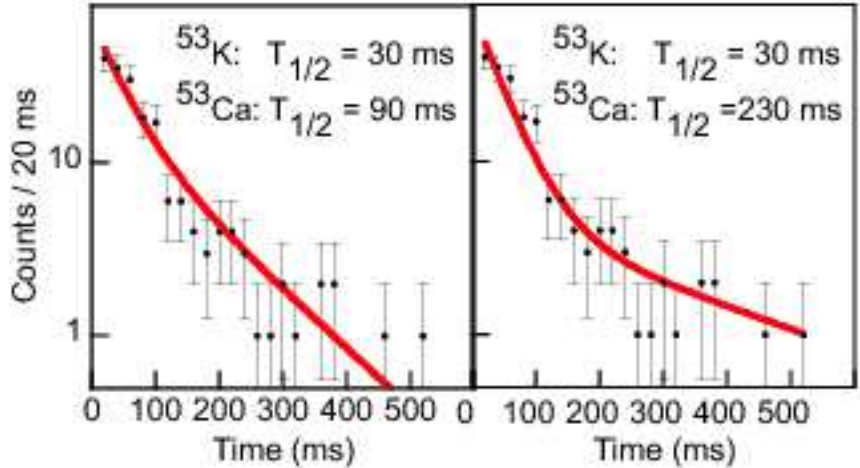}
\caption{(Color online) $\beta$-delayed neutron decay after collection
of $^{53}$K taken from Ref.\ \protect\cite{lan1983}.  Data were fitted with 
two different half-life values of the daughter $^{53}$Ca.  Regression analysis
results in $R^2$ values of 0.952 and 0.924 for the fits with $^{53}$Ca 
half-life values of 90~ms and 230~ms, respectively.}
\label{fig4}
\end{figure}

In the present work, the decay curve derived from $\beta$-delayed neutron
events, presented in Fig.\ 2 of Ref.\ \cite{lan1983}, was refitted with 
a function based on the Batemann equations, where the parent exponential
decay and daughter exponential growth and decay were considered.  The 
parent half-life value was fixed at $30$~ms, and two fixed values 
of the half-life of the $^{53}$Ca were used: the $90$-ms value deduced
by Langevin {\it et al.} \cite{lan1983} and the value of $230$~ms 
deduced here.  The parent activity was taken as a free parameter, and
the daughter activity was determined as 40\% of the parent activity,
based on the measured delayed-neutron branching following $\beta$ decay
of this nuclide \cite{lan1983}.   The resulting fits are compared with 
the data in Fig.\ \ref{fig4}.  An $R^2$ regression analysis to check the 
goodness of fit produced values of 0.952 and 0.924 for the functions that
used $^{53}$Ca half-life values of $90$~ms and $230$~ms, respectively.
The difference in $R^2$ values between the two fits is not significant,
if the errors in the half-life values of the parent (17\%) and daughter
($\sim$~20\%) are considered.  The longer half-life value
deduced here for $^{53}$Ca appears consistent with the $\beta$-delayed
neutron decay data obtained in the previous study of the $A = 53$ decay 
chain.

However, it should be acknowledged that 
we cannot rule out that the longer half-life measured
here could result from the presence of a second 
$\beta$-decaying state in 
$^{53}$Ca.  Isomeric $J^{\pi} = 9/2^+$ states are
known in the neutron-rich $^{61}$Fe$_{35}$ \cite{grz1998} 
and $^{59}$Cr$_{35}$ \cite{fre2004,grz1998}
isotopes.  Both the magnetic dipole moment \cite{mat2004} and 
electric quadrupole moment \cite{ver2007}
suggest the $9/2^+$ level in $^{61}$Fe is 
characterized by a deformed potential.  Although the neutron-rich Ti
and Ca isotopes should show less deformed structures
due to their proximity to the $Z=20$ shell closure, the possible
influence of the $\nu g_{9/2}$ orbital on the low-energy structure
of these isotopes requires further investigation.

\subsection{$^{54}$Ca}

\begin{figure}[h]
\includegraphics{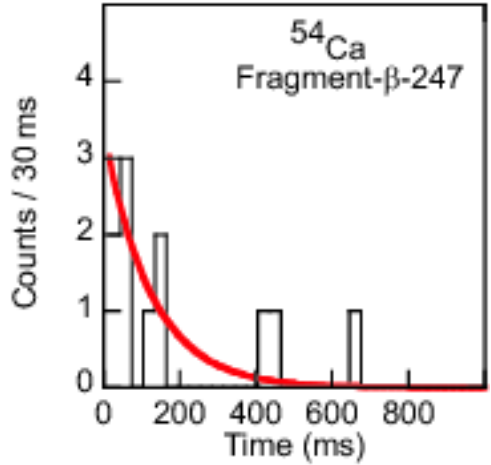}
\caption{(Color online) Decay curve for $^{54}$Ca showing fragment-$\beta$ correlations
with an additional requirement of a 247-keV $\gamma$ ray in coincidence.  The 
data were fitted with a single exponential decay with fixed parent half-life ($86$~ms) and 
the initial activity as a free parameter.}
\label{fig5}
\end{figure}

No previous information on the half-life of $^{54}$Ca was available.
The deduced half-life, $86 \pm 7$~ms, was obtained by fitting the 
fragment-$\beta$ decay curve in Fig.\ \ref{fig2} to a function 
that included exponential decay of the parent and exponential growth 
and decay of the $^{54}$Sc 
daughter (the constant background term was small and could be 
neglected).  The $^{54}$Sc half-life value was fixed at $360$~ms \cite{lid2004},
while the parent half-life and activity values were free parameters.
A single $\gamma$ ray with energy $246.9 \pm 0.4$~keV has been assigned to the 
decay of $^{54}$Ca based on the delayed $\gamma$-ray spectrum presented
in Fig.\ \ref{fig3}(b).  The decay curve constructed from $^{54}$Ca-$\beta$ 
correlations with an additional requirement of a coincidence with a 247-keV
$\gamma$ ray is given in Fig.\ \ref{fig5}.  The fitted decay curve 
returns a half-life consistent with
that deduced from the fragment-$\beta$ correlation data.     

\begin{figure}[h]
\includegraphics{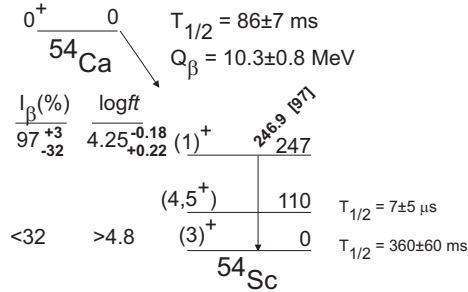}
\caption{Low-energy level structure of $^{54}$Sc and $\beta$-decay
scheme for the parent nuclide, $^{54}$Ca.  The energies of each state are
given in keV.  The level with energy 110~keV was previously identified
as a microsecond isomer by Grzywacz {\it et al.} \protect\cite{grz1998} and
later confirmed by Liddick {\it et al.} \protect\cite{lid2004}.  The 110-keV 
$\gamma$-ray was not observed in the $\beta$-decay study of $^{54}$Ca
presented here.}
\label{fig6}
\end{figure}

The absolute intensity of the 247-keV transition was determined to be 
$(97 \pm 32)$\%, based on a 
Gaussian fit to the $\gamma$-ray peak in Fig.\ \ref{fig3}(b), the measured
peak efficiency curve for the 16-detector $\gamma$-ray array, and 
the number of $^{54}$Ca implantations correlated with $\beta$ decay (see Table \ref{tab1}).
The absolute intensity implies that a majority of the 
apparent $\beta$ intensity from the decay of $^{54}$Ca proceeds through an excited
state depopulated by the 247-keV transition.  This transition is proposed to
directly feed the ground state of $^{54}$Sc.  The new 247-keV excited state 
is tentatively assigned spin and parity $1^+$, based on 
selection rules for allowed $\beta$ decay.    
The uncertainty in the spin and parity assignment 
for the 247-keV level arises from the large $Q$ value 
associated with the $^{54}$Ca $\beta$ decay.  The low statistics 
in the delayed $\gamma$-ray spectrum in Fig.\ \ref{fig3}(b) 
cannot exclude the presence of higher energy, lower intensity
transitions that may directly feed the level at 247 keV.     
The $\beta$-decay sequence and known levels in $^{54}$Sc are displayed
in Fig.\ \ref{fig6}.  The decay $Q$-value is taken from Ref.\ \cite{ame2003}.
The excited state at 110~keV in $^{54}$Sc was identified as a $\mu$s isomer 
in studies by Grzywacz {\it et al.} \cite{grz1998} and 
was reaffirmed in Ref.\ \cite{lid2004}.  The spin and parity of the 
$^{54}$Sc ground state was limited to $(3,4)^+$, based on the $\beta$-decay
feeding pattern to excited states in $^{54}$Ti \cite{jan2002}.  Placement 
of the 247-keV transition would preclude $J^{\pi} = 4^+$ for the 
final state of this``prompt'' $\gamma$-ray transition, therefore,
spin and parity $3^+$ is proposed for the ground state of $^{54}$Sc. 
The 247-keV transition is presumed to have $E2$ multipolarity, and 
the estimated half-life from Weisskopf estimates of 51~ns is consistent with
the ``prompt'' nature of this transition.

\subsection{$^{55}$Ca and $^{56}$Ca}

The only previous information regarding the isotopes $^{55,56}$Ca was 
their existence, established from fission of relativistic $^{238}$U ions
\cite{ber1997}.  In that study, six and three counts were associated
with the production of $^{55}$Ca and $^{56}$Ca, respectively.  We have achieved
sufficient statistics in the present study to deduce half-life values
for $^{55,56}$Ca for the first time.  The fragment-$\beta$ decay curves
were fitted to a function that included the exponential decay of the 
parent, the growth and decay of the daughter, and a linear background.  

\begin{figure}[h]
\includegraphics{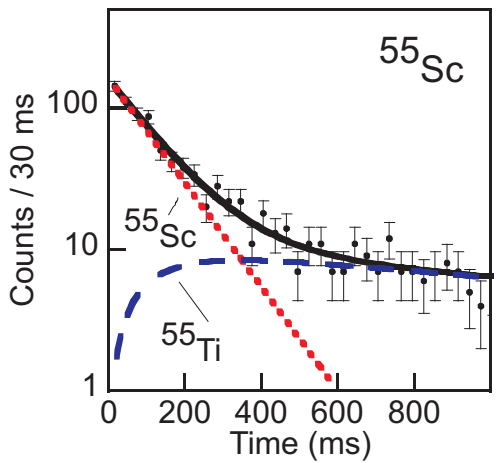}
\caption{(Color online) Decay curve for $\beta$ particles correlated with 
$^{55}$Sc implantation events.  The decay data were fitted
to a function (solid black line) that considered the exponential decay of the 
$^{55}$Sc parent (dotted line) and exponential growth and decay of the 
$^{55}$Ti daughter (dashed line) with known 
half-life $1.3 \pm 0.1$~s \protect\cite{man2003}.}
\label{fig7}
\end{figure}

For the $^{55}$Ca decay curve in Fig.\ \ref{fig2}(c), 
the $^{55}$Sc daughter half-life was fixed to a
value $83 \pm 3$~ms.  This new half-life was obtained 
in the present work from the analysis 
of $\beta$-decay events correlated with $^{55}$Sc implantations shown
in Fig.\ \ref{fig7}, and represents a 
significant improvement in precision with respect to previously
reported values \cite{sor1998,lid2004}.  The parent $^{55}$Ca half-life, initial
parent activity, and constant value for background were taken as 
free parameters.  The grand-daughter $^{55}$Ti has a known half-life of 
$1.3 \pm 0.1$~s \cite{man2003} and was not considered in the curve fitting function.  
The fit resulted in a half-life of $22 \pm 2$~ms for $^{55}$Ca 
(see Fig.\ \ref{fig2}), with a negligible background contribution.  

A single $\gamma$-ray transition of energy 592~keV was evident in the 
$\beta$-delayed $\gamma$-ray spectrum correlated with $^{55}$Ca 
implantations shown in Fig.\ \ref{fig3}.  This transition was previously
assigned to the decay of the $^{55}$Sc daughter \cite{lid2004}.

The fit to the $^{56}$Ca decay curve in Fig.\ \ref{fig2}(d) considered 
a fixed value of $35 \pm 5$~ms for the half-life of the $^{56}$Sc 
daughter.  There are, in fact, two $\beta$-decaying states in $^{56}$Sc
\cite{lid2004}.   Decay of the $^{56}$Ca ground state, with 
presumed spin and parity $0^+$, would populate 
lower-spin states in the daughter nucleus.  Therefore, the
dominant decay constant to consider for fitting
the $^{56}$Ca decay curve is that associated with the 
low-spin ($1^+$) level in $^{56}$Sc.  The half-life deduced
for $^{56}$Ca $\beta$ decay was $11 \pm 2$~ms, about a factor 
of two shorter than that in the neighbor $^{55}$Ca.  No evidence
was found for delayed $\gamma$-ray transitions correlated with
$^{56}$Ca $\beta$ decays.  However, the small number of 
correlated decays (18 total decay events -- see Table \ref{tab1}) 
combined with the
measured $\gamma$-ray efficiency curve would
preclude observation of more than one or two counts for  
$\gamma$ rays with energies at the maximum of the 
Ge array efficiency curve.        

\section{Discussion}

\subsection{Structure of $^{54}$Sc}

The $\beta$ decay of $^{54}$Ca apparently proceeds 
to a single state in $^{54}$Sc residing 247 keV above the ground state.  
This state has tentatively been assigned 
$1^+$ spin and parity.  This new 
information for $^{54}$Sc can be combined with 
previous knowledge on isomeric and $\beta$ decay 
to better understand the coupling 
of valence protons and neutrons in this odd-odd nucleus.

\begin{figure}[h]
\includegraphics{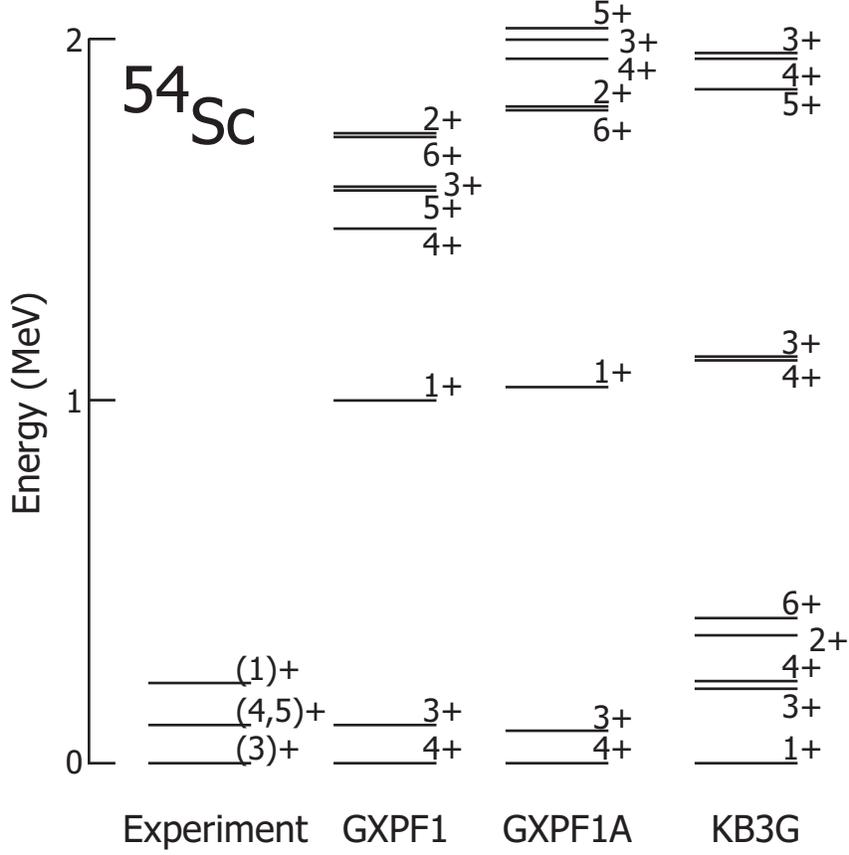}
\caption{Low-energy structure of $^{54}$Sc
compared to results of shell model calculations
employing the GXPF1 \protect\cite{hon2002,hon2004}, GXPF1A \protect\cite{hon2005},
and KB3G \protect\cite{pov2001} effective interactions.}
\label{fig8}
\end{figure}

The known low-energy levels in $^{54}$Sc in Fig.\ \ref{fig8} are compared
with results of shell model calculations completed in the 
full $pf$-shell model basis with the interactions GXPF1 \cite{hon2002,hon2004},
GXPF1A \cite{hon2005}, and KB3G \cite{pov2001}.  The difference in 
calculated level orderings between the 
GXPF1 and KB3G interactions
can be related to the relative energies of the 
neutron single-particle $f_{5/2}$ and $p_{1/2}$ orbitals.
The $3^+$, $4^+$ doublet near the ground state in $^{54}$Sc
in the calculations with the GXPF1 and GXPF1A interactions
arises from the coupling of the $\nu p_{1/2}$ and $\pi f_{7/2}$
orbitals.  This configuration appears lowest in energy in $^{54}$Sc
since the effective single-particle energy of the $\nu p_{1/2}$ orbital 
is below that of the $\nu f_{5/2}$ orbital with the GXPF1 and GXPF1A
interactions.  The first $1^+$ level is attributed mainly to
the coupling of the $\nu f_{5/2}$ and $\pi f_{7/2}$ single-particle
orbitals.  This level is at the ground state in the shell-model
results with the KB3G interaction, as this interaction
places the effective single-particle energy of the 
$\nu p_{1/2}$ orbital above that of the $\nu f_{5/2}$ orbital.
The position of the tentative $1^+$ level at 247~keV
assigned in this work, relative to the
$(3)^+$ ground state, is not well 
reproduced by either the KB3G, GXPF1, or GXPF1A interactions.   

The GXPF1 interaction proved successful in reproducing
the yrast structures of the Ti isotopes up to $^{56}$Ti$_{34}$,
where the first excited $2^+$ state was found to have 
an energy of 1129~keV \cite{lid2004-prl}, nearly 400~keV below
the GXPF1 expectation.  Modification of three
two-body pairing matrix elements associated with
the $f_{5/2}$, $f_{7/2}$, and $p_{1/2}$ orbitals  
and two quadrupole-quadrupole interaction 
matrix elements coupling the $f_{5/2}$ and $p_{1/2}$
orbitals to $J=2$ and $J=3$ resulted in the GXPF1A 
interaction \cite{hon2005}.  The shell model
results with the GXPF1A interaction  
reproduced the low-energy yrast structure of 
$^{56}$Ti \cite{for2004} better, while maintaining good agreement with 
the known level structures of the less neutron-rich Ti isotopes.  
As noted above, neither interaction does well in 
reproducing the position of the tentative $1^+$ state in 
$^{54}$Sc at 247~keV, which should be dominated by the shell model 
configuration $\pi f_{7/2}^1 \nu f_{5/2}^1$.
The low energy of the $(1)^+$ level relative to the 
proposed $(3)^+$ ground state, with dominant configuration
$\pi f_{7/2}^1 \nu f_{5/2}^1$, suggests that while the 
level ordering of the $p_{1/2}$ and $f_{5/2}$ neutron
single-particle orbitals in the GXPF1 and GXPF1A interactions
is likely correct, the effective energy spacing between
the neutron $p_{1/2}$ and $f_{5/2}$ orbitals is overestimated 
or the interaction strength between $\pi f_{7/2}^1$ and 
$\nu f_{5/2}^1$ is underestimated.  

Another outstanding question with regards to the low-energy
structure of $^{54}$Sc is the tentative spin-parity assignment
for the isomeric state located 110~keV above the ground
state.  Grzywacz {\it et al.} \cite{grz1998} deduced $E2$ multipolarity
for the 110~keV $\gamma$ ray, based on a comparison of
the experimental half-life value $(7 \pm 5)\mu$s 
with single-particle estimates.  Tentative spin and parity
$5^+$ was then assigned to the 110-keV excited state.
In Ref.\ \cite{lid2004}, it was suggested that 
the $E2$ isomer may represent the decay of the first excited 
$1^+$ state, whose energy may be lower than predicted 
by the shell model calculations utilizing the GXPF1 interaction.
Indeed, the $1^+_1$ state is lower than predicted, tentatively 
established in $^{54}$Sc at 247~keV.  Regardless of the interaction
employed, the first excited $5^+$ state, associated with the 
$\pi f_{7/2}^1 \nu f_{5/2}^1$ configuration, is expected at 
an energy in excess of 1.5~MeV (see Fig.\ \ref{fig8}).
The shell model results all suggest a doublet of levels with
$J^{\pi } = 3^+,4^+$ near the $^{54}$Sc ground state.  These
two levels belong to the odd-odd $\pi f_{7/2}^1 \nu p_{1/2}^1$ multiplet.  
The transition probabilities $B(M1;3^+ \rightarrow 4^+)$
and $B(E2;3^+ \rightarrow 4^+)$ calculated with the 
GXPF1 and GXPF1A interactions are listed in Table \ref{tab2}, along
with the calculated half-lives.  The $M1$ strength dominates,
but the calculated partial half-life was in the range of a few 
nanoseconds, much longer than the 9.3~ps half-life calculated 
from single-particle estimates.  The $M1$ hindrance factor
appears significant for this $\gamma$-ray transition.  Therefore,  
the long ($\mu$s) half-life established for the isomeric transition in 
$^{54}$Sc does not exclude possible $4^+$ spin and parity 
for the first-excited state.

\begin{table}[h]
\caption{Transition probabilities and partial half-lives for the 
$\gamma$-ray transition between the $3^+$ and $4^+$ members of the 
$\pi f_{7/2}^1 \nu p_{1/2}^1$ multiplet in $^{54}$Sc.  The experimental
transition energy of 110~keV was used in the calculations.} 
\label{tab2}
\begin{ruledtabular}
\begin{tabular}{ccccccc}
Multipolarity& GXPF1 & GXPF1A & Single-particle \\
$B(M1; 3^+ \rightarrow 4^+)$ $(\mu ^2)$   & 0.026  & 0.0037 &  \\
$T_{1/2}(M1; 3^+ \rightarrow 4^+)$ $(s)$  & $1.1 \times 10^{-9}$ 
& $7.9 \times 10^{-9}$ 
& $9.3 \times 10^{-12}$ \\
$B(E2; 3^+ \rightarrow 4^+)$ $(e^2 fm^4)$ & 4.50   & 5.55   & \\
$T_{1/2}(E2; 3^+ \rightarrow 4^+)$ $(s)$  & $5.7 \times 10^{-6}$ 
& $4.6 \times 10^{-6}$ 
& $2.9 \times 10^{-6}$  \\
\end{tabular}
\end{ruledtabular}
\end{table}

An alternative scenario for the 
low-energy structure of $^{54}$Sc is that the $3^+$
and $4^+$ states of the $\nu p_{1/2}$ $\pi f_{7/2}$ multiplet
are nearly degenerate in energy at the ground state.
The close energy spacing between these states would 
preclude a $\gamma$-ray transition between them.  The 
247-keV transition observed following the $\beta$
decay of $^{54}$Ca would then be an $E2$
transition between the $1^+$ and $3^+$ levels, while
the isomeric 110-keV $\gamma$ ray would also
have $E2$ multipolarity and would represent the 
transition between the $6^+$ and $4^+$ levels. 
Clearly, further studies will be required to provide
additional insight into the low-energy structure of $^{54}$Sc.

\subsection{$\beta$ decay half-lives of Ca isotopes}

\subsubsection{Gross Theory}

Gross $\beta$-decay theory provides a means for predicting $\beta$-decay
properties of nuclides far from the valley of stability.  The $\beta$-decay
half-lives are inversely proportional to the Fermi function, and hence, the 
$\beta$-decay $Q$ value raised to the fifth power: $T_{1/2} \sim Q_{\beta}^{-5}$.
In practice, half-life values calculated from gross theory are too long, as the 
decay strength to low-energy states is underestimated \cite{pfe2002}.  
Tachibana {\it et al.} \cite{tac1990} extended traditional gross $\beta$-decay
theory to include a one-particle strength function for the Gamow-Teller decay.
Pfeiffer, Kratz, and M\"{o}ller \cite{pfe2000} took a more phenomenological approach to
address missing $\beta$ strength to low-energy states by relating the 
half-life and $Q$ value in a manner similar to that employed by 
Kratz and Herrmann \cite{kra1973} to extract $P_{n}$ values:

\begin{equation}
T_{1/2} \sim a \times (Q_{\beta} - C)^{-b}
\label{eq1}
\end{equation} 

\begin{figure}[h]
\includegraphics{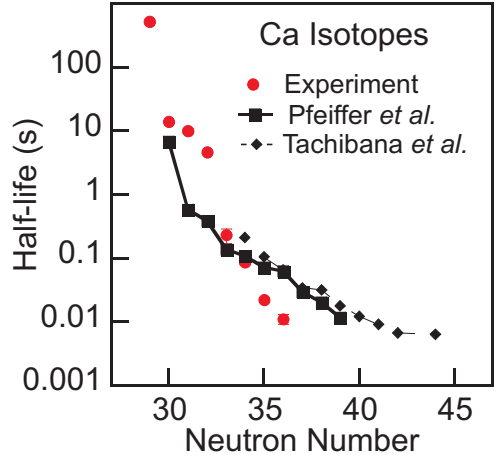}
\caption{(Color online) Experimental half-life values of neutron-rich
Ca isotopes (filled circles) compared with predictions of 
gross $\beta$-decay theory extended by Tachibana {\it et al.} \protect\cite{tac1990}
and Pfeiffer {\it et al.} \protect\cite{pfe2000}.}
\label{fig9}
\end{figure}

Here, $a = 2740$~s and $b = 4.5$ are taken as constants, 
and $C$ is a cutoff energy related to
the pairing gap in the daughter nucleus \cite{mad1988}.
The results of the two extensions to gross theory for the neutron-rich 
Ca isotopes are compared with experiment in Fig.\ \ref{fig9}.  The 
two approaches predict similar half-life values in the range 
$^{54-59}$Ca.  The calculations compare favorably with the experimental
half-lives of $^{53,54}$Ca.  Both predict too long 
half-lives for the more neutron-rich $^{55,56}$Ca isotopes, a continuing 
issue with gross $\beta$-decay theory, especially far from the valley 
of stability, where $Q_{\beta}$ is large.  The other apparent
deficiency in the calculations occurs at $^{51,52}$Ca, as the 
calculated short half-lives do not consider the known subshell closure
at $N=32$ for Ca.

\subsubsection{QRPA results and the fast $\beta$ decays of $^{54-56}$Ca }

\begin{figure}[h]
\includegraphics{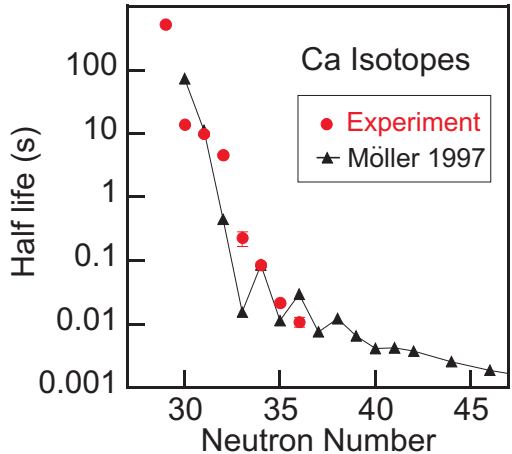}
\caption{(Color online) Experimental half-life values of neutron-rich
Ca isotopes (filled circles) compared with predictions of the QRPA compilation  
of M\"{o}ller {\it et al.} \protect\cite{mol1997}.}
\label{fig10}
\end{figure}

M\"{o}ller, Nix and Kratz \cite{mol1997} compiled an extensive database 
of ground-state properties of nuclides across the nuclear chart, including
$\beta$-decay properties.  Nuclear deformations
were based on the Finite-Range Droplet Model \cite{mol1995}, and Gamow-Teller
$\beta$-decay rates were calculated in the framework of the quasi-particle
Random Phase Approximation (QRPA).  The half-lives from the QRPA calculations 
are compared with experiment for the neutron-rich Ca isotopes in Fig.\ \ref{fig10}.
M\"{o}ller {\it et al.} \cite{mol2003} extended the QRPA compilations reported
in Ref.\ \cite{mol1997} to include first-forbidden $\beta$ decays.  Such decays
should, in general, be more important for nuclides nearer stability, where $Q_{\beta}$
values are small.  Comparison of the previously calculated half-lives of the neutron-rich 
Ca isotopes with this new approach showed little difference, therefore, only
the QRPA results for the Gamow-Teller decay rates are given in Fig.\ \ref{fig10}. 

The trend in calculated half-lives follows well the 
experimental values over the range $^{50-56}$Ca.  Again, the slow rate of 
the $^{52}$Ca $\beta$ decay is indicative of the subshell gap at 
$N = 32$.  The strong odd-even effect apparent in the QRPA calculations
is not manifested in the experimental half-lives.  Indeed, it is only 
in $^{52}$Ca that a deviation occurs from the regularly decreasing half-lives in
the Ca isotopes.  That $^{54}$Ca$_{34}$ follows this regular behavior and 
shows a (relatively) fast $\beta$ decay may indicate that an appreciable  subshell gap
at $N=34$ has not developed for the Ca isotopes.

The fast $\beta$ decays observed experimentally in $^{55,56}$Ca reflect the 
spin-flip nature of these processes.  The last neutron(s) in
 $^{55}$Ca$_{35}$ and $^{56}$Ca$_{36}$ should occupy the 
$f_{5/2}$ neutron orbital.  $\beta ^-$ decay is expected to proceed as
the spin-flip decay $\nu f_{5/2} \rightarrow \pi f_{7/2}$.  Such fast decays
should have small log$ft$ values, and the most prominent $\beta$-decay
branch should be to the ground state of the daughter.  Expectations 
regarding the $\beta$-decay branching in the neutron-rich Ca isotopes
are discussed in the next section.  

\subsubsection{Shell Model results with GXPF1}

\begin{figure}[h]
\includegraphics{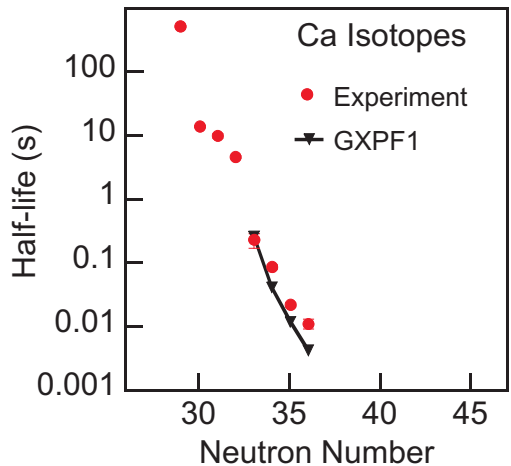}
\caption{(Color online) Experimental half-life values of neutron-rich
Ca isotopes (filled circles) compared with results of shell model calculations
in the full $pf$-model space with the GXPF1 interaction.}
\label{fig11}
\end{figure}

The $\beta$-decay properties of the $^{53-56}$Ca isotopes were calculated 
with the shell model effective interaction GXPF1 \cite{hon2002,hon2004} 
in the full $pf$ model space with the 
{\sc CMICHSM} \cite{hor2003} code. 
$\beta$-decay $Q$ values were taken from Ref.\ \cite{ame2003},
and the Gamow-Teller rates were quenched by a factor of 0.77, 
as explained in Ref.\ \cite{pov2001}.  The calculated half-lives 
agree within a factor of two with experimental values as demonstrated in Fig.\ \ref{fig11}.
A summary of the shell model results is also given in Table \ref{tab3}. 

\begin{table}[h]
\caption{Shell model results with the effective interaction GXPF1 for 
the $\beta$-decay properties of $^{54-56}$Ca.  Experimental values were obtained
in this work, unless otherwise indicated.} 
\label{tab3}
\begin{ruledtabular}
\begin{tabular}{ccccccc}
Nuclide& $J^{\pi}$&$T_{1/2}^{calc.}$ (ms)&$T_{1/2}^{exp.}$ (ms)&$P_n^{calc.}$ (\%)
              &$P_n^{exp.}$ (\%)& calc.\ g.s.\ Branch (\%)\\
$^{53}$Ca  &$1/2^-$ & 264   &  $230 \pm 60$ & 6 & $40 \pm 10$ \footnotemark[1] &  0 \\
$^{54}$Ca  &$0^+$   &  41   &  $ 86 \pm  7$ & 6 & $< 32$                       &  0 \\
$^{55}$Ca  &$5/2^-$ &  11.7 &  $ 22 \pm  2$ & 4 &                              & 92 \\
$^{56}$Ca  &$0^+$   &   4.3 &  $ 11 \pm  2$ & 2 &                              & 97 \\
\end{tabular}
\end{ruledtabular}
\footnotetext[1]{Ref.\ \protect\cite{lan1983}}.
\end{table}

The $\beta$ decay of $^{53}$Ca from the presumed $J^{\pi} = 1/2^-$ ground state 
is predicted to feed the first excited $3/2^-$ state in $^{53}$Sc with intensity
60\% of the total feeding strength.  The $\beta$ strength to neutron unbound 
states is calculated to be 6\%, which is smaller than that observed in Ref.\ \cite{lan1983}.
The expected decay of $^{54}$Ca is similar to that of $^{53}$Ca.  About 60\%
of the calculated $\beta$-decay strength is predicted to feed the first $1^+$ state
in $^{54}$Sc, and the $P_{n}$ value is also calculated to be 6\%.  The observed
$\beta$-decay strength from $^{54}$Ca is apparently dominated by decay to the 
the 247-keV level in $^{54}$Sc.  The absolute intensity of the 247-keV $\gamma$-ray
transition that depopulates this state is 97\%, with no other 
peak (intensity threshold $< 10\%$ below 1~MeV) in the delayed $\gamma$-ray spectrum
displayed in Fig.\ \ref{fig3}(b).  However, as noted earlier, only tentative
$1^+$ spin and parity have been proposed for this 247-keV level, since the low
statistics of the measurement does not rule out the presence of higher-energy
$\gamma$ rays feeding the 247-keV level from potential higher-lying $1^+$ states.
The shell model results with the GXPF1 interaction support the notion
that the primary branching in the $\beta$ decay of $^{54}$Ca is to the 
first excited $1^+$ in $^{54}$Sc.   
The experimental delayed-neutron branching reported 
in Table \ref{tab3} for
$^{54}$Ca is governed by the lower limit of the absolute intensity of the 
247-keV $\gamma$ ray.  

The $\beta$-decay properties of $^{55,56}$Ca 
were found to be similar to the shell model results, as was 
the case with $^{53,54}$Ca.  The decays are 
dominated by fast ground state to ground state transitions, which carry more 
than 90\% of the $\beta$-decay strength.  Since only a small number of decays
were detected for both $^{55}$Ca and $^{56}$Ca, there was little possibility
to observe branching to excited states in the $^{55}$Sc and $^{56}$Sc
daughters, respectively, that would lead to $\gamma$-ray or neutron emission.
The short half-life values and large, direct ground-state $\beta$ branching 
again reflect the dominance of the spin-flip $\nu f_{5/2} \rightarrow \pi f_{7/2}$
decay process in the shell model results.  The good agreement between the 
experimental and shell model half-lives for $^{55,56}$Ca support the 
notion that the $\nu f_{5/2}$ orbital is near the Fermi 
surface for these nuclides with $N>34$.

\section{Summary}

The $\beta$-decay properties of the neutron-rich Ca isotopes have 
been extended to $^{56}$Ca$_{36}$.  
The half-lives of
$^{54-56}$Ca have been measured for the first time to be 
$86 \pm 7$~ms, $22 \pm 2$~ms, and $11 \pm 2$~ms, respectively.
A new half-life of 
$230 \pm 60$~ms has been deduced for $^{53}$Ca, a value
longer than that reported previously.  
The trend of the $\beta$-decay half-lives of the 
neutron-rich Ca isotopes shows a regular decrease except at 
$^{52}$Ca$_{32}$, which is attributed to the known subshell 
closure at $N=32$.  The low-energy structure of $^{54}$Sc, the 
$\beta$-decay daughter of $^{54}$Ca, has been elucidated further.
The ground and first excited states most likely arise from
coupling of the $\pi f_{7/2}$ and $\nu p_{1/2}$ orbitals, suggesting
that the $\nu p_{1/2}$ single-particle orbital resides below
the $\nu f_{5/2}$ state at $N=33$.  Such level
ordering was also inferred from the low-energy
structure of the $N=33$ isotone $^{55}$Ti \cite{zhu2007}.  
The first $1^+$ level in $^{54}$Sc
has been proposed at an energy of 247 keV above the ground state.  
This tentative $1^+$
level is presumed to be a member of the $\pi f_{7/2} - \nu f_{5/2}$ 
multiplet, and is nearly 800~keV lower than shell model predictions
with the GXPF1 and GXPF1A effective interactions.  It seems
that the effective energy gap between adjacent neutron 
single-particle orbitals $f_{5/2}$
and $p_{1/2}$ is overestimated or the 
interaction strength between 
the $\pi f_{7/2}$ and $\nu f_{5/2}$
orbitals is underestimated by the GXPF1 interaction.
The fast $\beta$ decays of $^{55,56}$Ca agree within a factor 
of two with the shell model results, where the $\beta$ strength
is dominated by the $\nu f_{5/2} \rightarrow \pi f_{7/2}$ 
spin-flip decay.  The shell model prediction for the 
half-life of $^{54}$Ca is also 
within a factor of two of experiment.  Such good agreement
may initially invoke support for a shell 
closure at $N=34$, since the GXPF1 interaction
includes a significant energy gap between 
adjacent single-particle orbitals $\nu f_{5/2}$ and $\nu p_{1/2}$.  
However, the $^{54}$Ca half-life is also found to agree well with the 
predictions of gross $\beta$ decay theory and the results
of QRPA calculations, which do not formally consider a strong
shell gap at $N=34$. Therefore, any interpretation regarding
magicity at $^{54}$Ca$_{34}$ will require further experimental 
investigation.  Exploring the $\beta$-decay properties 
of heavier Ca isotopes towards $^{60}$Ca will prove challenging 
due to low production cross sections. However, such studies 
are vital to better understand the evolution of neutron 
single-particle states with an increase in isospin 
and corresponding decrease in neutron separation energy. 

\begin{acknowledgments}

The authors thank the NSCL operations
staff for providing the primary and
secondary beams for this experiment and members
of the NSCL Gamma group for assistance in 
setting up and maintaining the Ge detectors.  
The work was supported in
part by the National Science Foundation grants
PHY-06-06007 (NSCL), PHY-02-44453 (JINA),
PHY-04-56463 (FSU), and PHY-05-55366 (CMU/MSU)
and the U.S.\ Department of 
Energy, Office of Nuclear Physics, under 
contracts DE-AC02-06CH11357 (ANL) and 
DE-FG02-94-ER40834 (Maryland).  Travel support
was also provided by the Polish Science Committee
grant P03B 059 29.  
 
\end{acknowledgments}

\end{document}